# Transport evidence for the coexistence of the topological surface state and a two-dimensional electron gas in BiSbTe$_3$ topological insulator

Fei-Xiang Xiang, Xiao-Lin Wang*, and Shi-Xue Dou

*Spintronic and Electronic Materials Group, Institute for Superconducting and Electronic Materials, Australian Institute for Innovative Materials, University of Wollongong, North Wollongong, 2500, Australia*

*(April 30, 2014)*

## Abstract

Topological insulators (TIs) are new insulating materials with exotic surface states, where the motion of charge carriers is described by the Dirac equations and their spins are locked in a perpendicular direction to their momentum. Recent studies by angle-resolved photoemission spectroscopy have demonstrated that a conventional two-dimensional electron gas can coexist with the topological surface state due to the quantum confinement effect. The coexistence is expected to give rise to exotic transport properties, which, however, have not been explored so far. Here, we report a magneto-transport study on single crystals of the topological insulator BiSbTe$_3$. Besides Shubnikov-de Haas oscillations and weak anti-localization (WAL) from the topological surface state, we also observed a crossover from the weak anti-localization to weak localization (WL) with increasing magnetic field, which is temperature dependent and exhibits two-dimensional features. The crossover is proposed to be the transport manifestation of the coexistence of the topological surface state and two-dimensional electron gas on the surface of TIs.

Topological insulators (TIs) are new states of quantum matters characterized by a bulk gap, but gapless edge or surface states for two-dimensional (2D) or three-dimensional (3D) TIs, which host helical Dirac fermions[1-3]. These new states originate from the strong spin-orbit interaction of heavy elements, which causes inversion of the valence and conduction bands and therefore, generates a gap in the bulk, while there is a gapless state on the surface that is protected by time-reversal symmetry and robust against non-magnetic perturbations. The exotic topological surface state provides the opportunity to study the novel quantum Hall effect[4] and Majorana fermions[5], and promises applications in magneto-electric sensors[6], spintronics and quantum computation[1-3].

The topological surface state has been verified by surface sensitive techniques, such as angle-resolved photoemission spectroscopy (ARPES) and scanning tunnelling microscopy (STM) and spectroscopy (STS)[7-10]. It has also been revealed in various quantum phenomena by transport measurements, such as Aharonov-Bohm (AB) oscillations, Shubnikov-de Haas (SdH) oscillations, and weak anti-localization (WAL)[11-14]. Very recently, an inversion asymmetric topological insulator (IATI), BiTeCl, was discovered by ARPES[15]. The IATI not only exhibits new topological phenomena such as crystalline-surface-dependent topological electronic states, pyroelectricity, and an intrinsic topological $p$-$n$ junction, but also provides an ideal platform for the realization of topological magneto-electric effects[15]. The topological surface state in BiTeCl has been demonstrated by the observation of SdH oscillations, which exhibit the 2D nature of the Fermi surface and $\pi$ Berry phase[16]. In addition, when the Fermi level is close to the Dirac point, the topological surface electrons exhibit a small effective mass, $0.055 m_e$, and quite large mobility, $4490$ cm$^2$s$^{-1}$ [16].

We note that recent ARPES investigations have revealed strong evidence that a two-dimensional electron gas (2DEG) and the topological surface state (TSS) can coexist on the surface of topological insulators such as $Bi_2Se_3$ and $Bi_2Se_{3-x}Te_x$[17-24]. It has been proposed that 2DEG is caused by the formation of quantum wells due to surface perturbations, which is believed to be generally present in narrow-gap topological insulators. The modification of the topological electronic structure and the formation of the additional quantum well states are expected to give rise to new phenomena. For example, by tuning the electric field gradient of the quantum wells, the 2DEG can exhibit large tunable Rashba spin splitting, which promises applications in non-magnetic spintronics[17,20,25].

In a transport study of topological insulators, the SdH oscillations of TSS exhibit 2D characteristics due to the 2D Fermi surface and the linear fit of the Landau level index in the Landau level fan diagram intercepts, with the $n$-index axis at ±0.5 due to the π Berry phase. The π Berry phase will also lead to weak anti-localization (WAL), as a result of the destructive interference of the electronic wave, while the 2DEG would show either WAL or weak localisation (WL), depending on the phase coherence of the electronic waves. According to the ARPES observations, the 2DEG sub-bands do not split, which is indicative of little Rashba effect. The coexistence can lead to both the WAL and WL from the TSS and 2DEG. Furthermore, a crossover from WAL to WL can be expected, depending on the number of 2DEG channels. Little work has been done, however, on the transport properties associated with the coexistence of the topological surface state and the 2DEG so far.

In this work, by performing transport measurements on $BiSbTe_3$ single crystal with the Fermi level located in the bulk band gap, we detected the Shubnikov-de Haas oscillations and weak anti-localization from the topological surface state (TSS) with large carrier mobility. Moreover, we observed a crossover from weak anti-localization to weak localization, which is possibly the transport manifestation of the coexistence of the TSS and conventional 2DEG.

The challenge for transport studies is the bulk carrier contribution to the transport, which hinders the detection of the topological surface state. Gating, chemical doping, more careful growth of the materials, and nanosized material are the frequently used methods to suppress the bulk contribution[14,26-28]. In particular, both $Bi_2Te_3$ and $Sb_2Te_3$ are 3D topological insulators, and it has been verified that, similar to the III-V semiconductor $Al_xGa_{1-x}As$, the isostructural and isovalent alloy of $Bi_2Te_3$ and $Sb_2Te_3$ can effectively engineer the bulk properties while maintaining the topological surface state[27,29]. By band engineering in $(Bi_{1-x}Sb_x)_2Te_3$ thin film and nanoplates, the chemical potential has been systematically tuned whenever the topological surface state still exists[27,29]. The carrier mobility of the topological surface state of $(Bi_{1-x}Sb_x)_2Te_3$ is so small, however, that the quantum oscillation from the surface state of this material is hard to observe[29,30]. Therefore, it is quite necessary to grow high quality single crystal with large carrier mobility, which is important for the study of its electronic properties by transport measurements. In this series of compounds, the Fermi level of $BiSbTe_3$ single crystal is located in the bulk gap, which is ideal for transport investigation, and we therefore chose $BiSbTe_3$ single crystal for our study.

Figure 1(a) shows the temperature dependence of the resistance of cleaved single crystal $BiSbTe_3$ in magnetic field of 0, 5, and 13 T. From 300 K to 155 K, the resistance decreases with temperature, exhibiting metallic behaviour. The resistance increases with temperature below 150 K, exhibiting insulating behaviour, which means that the insulating bulk state is achieved. The activation energy deduced from the linear part of Arrhenius plot of $\ln\rho$ vs. $1/T$, as shown in the inset of Fig. 1(a), is 26.9 meV. It should be noted that there is a crossover between zero field and non-zero field *R-T* curves, which is an indication of negative magnetoresistance and will be discussed in detail below. Fig. 1(b) shows the typical Hall resistance at 2.5 K for a $BiSbTe_3$ single crystal from the same batch, which indicates that the dominant carriers are electrons. To determine the relative position of the Fermi level in the band structure, the magnetic field dependence of the Hall resistance, $R_H$, of a sample cleaved from same single crystal was measured, as shown in Fig. 1(b). The negative value of $R_H$ in positive field and the non-linear magnetic field dependence of $R_H$ indicate that at least two types of charge carriers contribute to $R_H$, but that the dominant charge carriers are electron. So, according to our experimental data and the ARPES results on the thin film samples, the band structure of our sample could be schematically sketched, as shown in Fig. 1(c).

To detect the TSS by the Shubnikov-de Haas oscillations, the magnetoresistance was measured in magnetic field up to ±13 T. The SdH oscillation can be seen clearly at 2.5 K in Fig.2(a) and gradually disappears with increasing temperature. Because the sample is already in the insulating regime, the oscillation of the magnetoresistance was inferred to be due to the electrons in the surface state. To verify the origin of the SdH oscillations, the angular dependence of the magnetoresistance was measured every 10º, as shown in Fig. 2(b), and the experimental set-up is shown in the inset of Fig. 2(b). In an electronic system with a two-dimensional (2D) Fermi surface (FS), the peak positions of the SdH oscillation depend only on the perpendicular component of the magnetic field $B_\perp = B\cos\theta$ to the sample surface. It can be seen that the oscillation disappears as the angle increases from 0° to 90°, suggesting the 2D nature of the FS. To clearly confirm the 2D nature of the SdH oscillation, $dR/dB$ vs. $1/B_\perp$ is plotted as shown in Fig. 2(c), in which the positions of its maxima and minima depend only on the perpendicular component of the magnetic field, $B_\perp$, indicating that the SdH oscillations have a 2D character.

The SdH oscillation originates from the Landau quantization of the FS, and the oscillation frequency is related to the extremal cross section of the FS, $S_F$, which is quantitatively described by the Onsager relation: $F = (\hbar c/2\pi e)S_F$, where $S_F = \pi k_F^2$, where $k_F$ is the Fermi wave vector. Through the fast Fourier transform, the oscillation frequency, $F = 50\ T$, is obtained; with this frequency, the Landau level fan is plotted, and the intercept is set to be 0.5. As shown in Fig. 2(d), all of the points are nearly located on the fitting line, which verifies the π Berry phase of the surface state. Thus, it could be concluded that the SdH oscillations in our sample originate from the TI surface state.

After verifying the origin of the SdH oscillations, the temperature and magnetic field dependence of the oscillation amplitude were analysed to extract the information on the surface states. The oscillation amplitudes were obtained after subtracting the smooth background of magnetoresistance (MR). The temperature and magnetic field can affect the resolution of the Landau tubes: the lower the temperature and the higher the magnetic field is, the larger the oscillation amplitude will be. This is well described by the Lifsitz-Kosevich theory, in which the thermal damping factor $R_T$ and the Dingle damping factor $R_D$ describe the temperature and magnetic field dependence of the oscillation amplitudes as follows: $R_T = aTm^*/B\sinh(aTm^*/B)$ and $R_D = \exp(-aT_D m^*/B)$ where $\sinh$ is the hyperbolic function, the effective mass $m = m^*m_e$, and $a = 2\pi^2 k_B/e\hbar \approx 14.69$ T/K, and $T_D = \hbar/2\pi k_B \tau$ is the Dingle temperature, $\tau$ the scattering time.

Fitting the temperature dependence of the oscillation amplitude with the thermal damping factor yields the effective mass $m_{cyc} = 0.105\ m_e$. From $m_{cyc}$ and $k_F$ we obtain the Fermi velocity $v_F = \hbar k_F/m_{cyc} = 4.23 \times 10^7$ cms$^{-1}$. Because $\Delta R/R_0 \sim R_D R_T$ (where R is the resistance), the Dingle temperature, 25 K, and the scattering time $\tau$, $4.87 \times 10^{-14}$ s, are obtained from the slope in the semilog plot of $\Delta RB\sinh(aTm^*/B)$ versus $1/B$, as shown in Fig. 2(d). This gives the mean free path $l_s = v_F \tau$ of 20.62 nm, where $v_F$ is the Fermi velocity, and the mobility $\mu = el_s/\hbar k_F \approx 821.95$ cm$^2$V$^{-1}$s$^{-1}$.

Besides the SdH oscillation at high field, another piece of evidence for the presence of a TSS in our sample is the weak anti-localization (WAL) in low field (Fig. 2(a)). The WAL emerges as the correction of coherent time-reversed closed paths to electronic transport when the phase coherent length, $l_\phi$, of electrons is much greater than the elastic scattering length, $l_e$. Applying the magnetic field will break the phase coherence and therefore results in a sharp increase (cusplike MR) of the resistance in magnetic field, which is the key signature of WAL. In topological insulators, the WAL is ascribed to the surface state with π Berry phase, which causes the destructive interference. As shown in Fig. 2(a) the weak anti-localization is pronounced at low temperatures such as 2.5 K and gradually disappears with increasing temperature, since the phonon vibrations increase with temperature, which degrades the phase coherence of the electrons. The quantum correction to the Drude conductivity is well described by the Hikami-Larkin-Nagaoka (HLN) formula[31], as follows:

$$\Delta\sigma_{HLN}(B) = \alpha \frac{e^2}{\pi h}\left[\Psi\left(\frac{\hbar}{4eBl_\phi^2}+\frac{1}{2}\right) - \ln\left(\frac{\hbar}{4eBl_\phi^2}\right)\right], \qquad (1)$$

Where $\psi$ is the digamma function, $B$ is the magnetic field, $l_\phi$ the phase coherence length, and α is the prefactor, while $l_\phi$ and α are fitting parameters. In our experiment, the low temperature magnetoresistance can be well fitted by this formula. As shown in Fig. 3, the low field magneto-conductance at 2.5 and 5 K is fitted well by the HLN formula, where α is -1.1 and -1.1, and $l_\phi$ is 40 and 31 nm, respectively. The prefactor, α = -1.1, in our sample indicates that both the bottom and top TSS exist.

Besides the ubiquitously observed WAL feature for topological insulators, a negative MR related to the weak localization (WL) is also observed in our measurements, as discussed below. As shown in Fig. 4(a, b), the WAL crosses to the negative MR at a field of around 1–2 T, which is clear at a low temperature such as 2.5 K, but with increasing temperature, the crossover become less pronounced and eventually disappears. At a high temperature such as 300 K, besides the disappearance of the WAL, the negative MR is transformed to positive MR, and the whole MR curve exhibits a parabolic shape corresponding to the MR caused by orbital scattering. Such temperature dependent negative MR is the signature of the WL effect. The theoretical calculations reveal that random magnetic scattering could drive the system from the symplectic to the unitary class, and if the magnetic doping can open a gap at the Dirac point, the weak anti-localization could be made to cross over to weak localization by tuning the gap or Fermi level[32]. The ARPES experiments have revealed a gap that opens at the Dirac point in magnetically doped topological insulators[33,34] and a crossover from weak anti-localization to weak localization when they are magnetically doped or in proximity to ferromagnetic thin films[34,35]. In addition, WL can also emerge in the bulk quantization regime whenever the bulk WL channels outnumber the surface WAL channels[36-38] Because the surface state bands and the lowest 2D bulk bands of a topological insulator thin film can be described by the two-dimensional modified Dirac model, a finite gap leads to the weak localization or the unitary behaviour. Neither of them, however is the reason for the behaviour of our non-magnetic bulk sample. The angular dependence of the MR shown in Fig. 2(b), however, gives us a clue as to the origin of the WL, where the crossover become less obvious with increasing angle and disappears from 50°, indicating the 2D nature of the negative MR. On the other hand, it has been reported that the 2D electron gas is present on the topological surface state when the surface is exposed to atmosphere, which is believed to be generally true for the topological insulators. The band structure of the 2D electron gas is similar to that of bulk quantized samples[17,20,25]. Because our transport measurements were performed under ambient atmosphere, a 2DEG can form on the surface of our sample, which will lead to the WL. Thus, the crossover from WAL to WL in our sample should arise from the coexistence of the 2DEG and TSS on the surface.

In conclusion, we studied the magneto-transport properties of single crystals of the topological insulator $BiSbTe_3$ with the Fermi level located in the bulk band gap. Shubnikov-de Haas oscillation was observed and identified as originating from the surface state by the angular dependence of the MR and π Berry phase. The high mobility, 821.95 cm$^2$V$^{-1}$s$^{-1}$, explains why SdH oscillation was observed. In addition to the WAL from the TSS, a crossover from weak anti-localization to weak localization was observed; we propose that the crossover should be ascribed to the coexistence of the topological state and a two-dimensional electron gas on the surface.

**Methods**

Single crystals of $BiSbTe_3$ were grown by slowly cooling a melt of the high purity (5N) elements Bi, Sb, and Te. The mixture of three elements with molar ratio of 1:1:3 was put into a quartz tube, which was evacuated and sealed, and then heated up to 800 ºC and held at that temperature for 24 h. It was then cooled to 550 ºC over several days, annealed at that temperature for 3 days, and finally furnace cooled to room temperature. From the X-ray diffraction, only 00l reflections appear with sharp peaks, indicating that the *c*-axis of the crystal is perpendicular to the cleavage plane of the single crystal. For transport measurements, the single crystals were cleaved and cut into a rectangular shape. The dimensions of the sample used in this study, except for the Hall measurements, are $0.58 \times 0.85 \times 0.09$ mm$^3$ (L × W × T, where W and T are the width and thickness respectively, and L is the distance between two voltage contacts). The standard four probe method was employed, and the Ohmic contacts were made with silver epoxy cured at room temperature. For the Hall resistance measurements, the standard six probe method was employed with the rectangular Hall bar shape. The magnetoresistance and Hall resistance were measured by sweeping the perpendicular magnetic field between ±13 T, except for the angular dependence of the magnetoresistance.

*Acknowledgements*
X.L.W. acknowledges the support from the Australian Research Council (ARC) through an ARC Discovery Project (DP130102956) and an ARC Professorial Future Fellowship project (FT130100778).

*Correspondence and requests for samples should be addressed to X.L.Wang (xiaolin@uow.edu.au).*

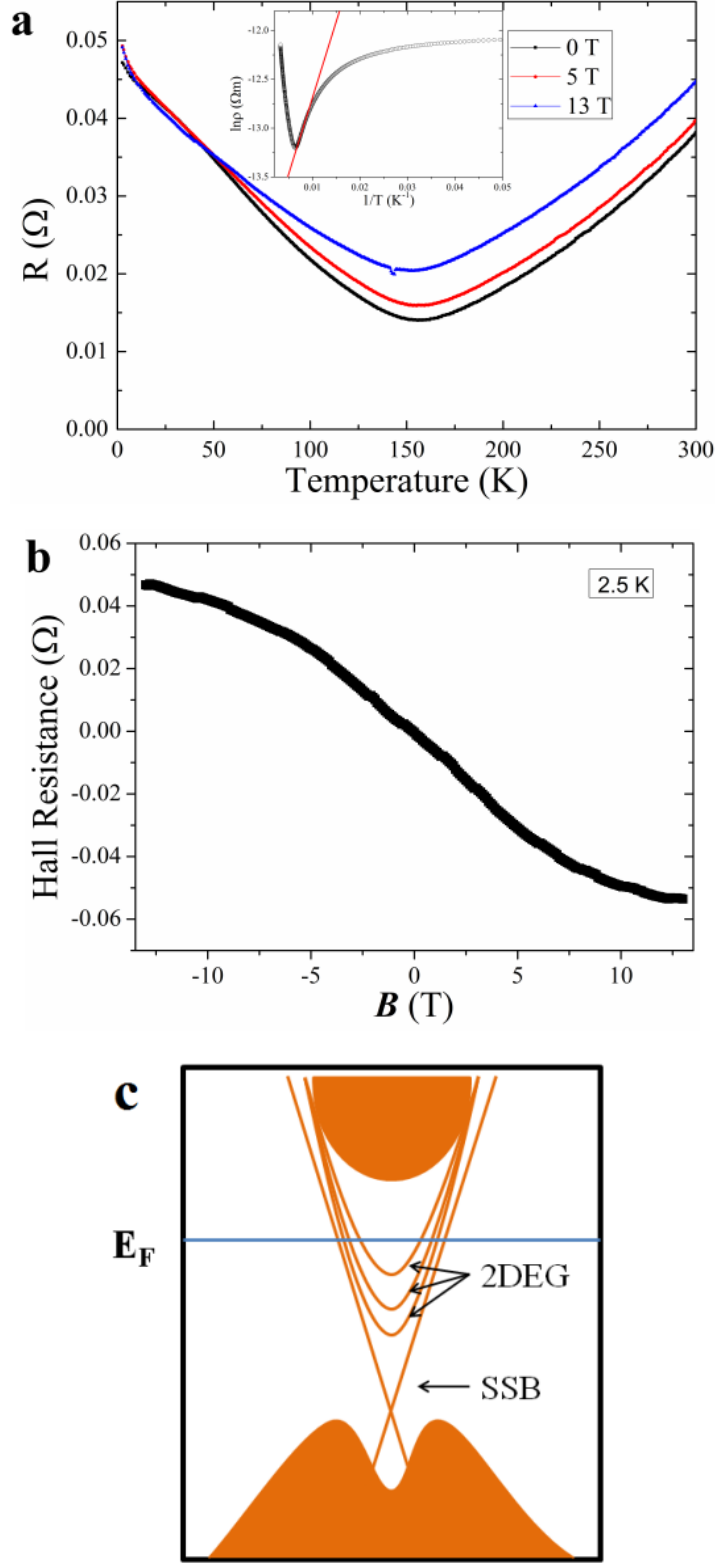

**Fig. 1** (a) Temperature dependence of the resistance at 0, 5, and 13 T. Below around 150 K, the resistance shows insulating temperature dependence. The inset contains the corresponding Arrhenius plot. (b) Hall resistance of the single crystal measured at 2.5 K. (c) Schematic diagram of the band structure for BiSbTe$_3$ single crystal. The Fermi level is located in the bulk band gap but close to the conduction band, and the 2DEG subbands are inferred from the crossover from WAL to WL.

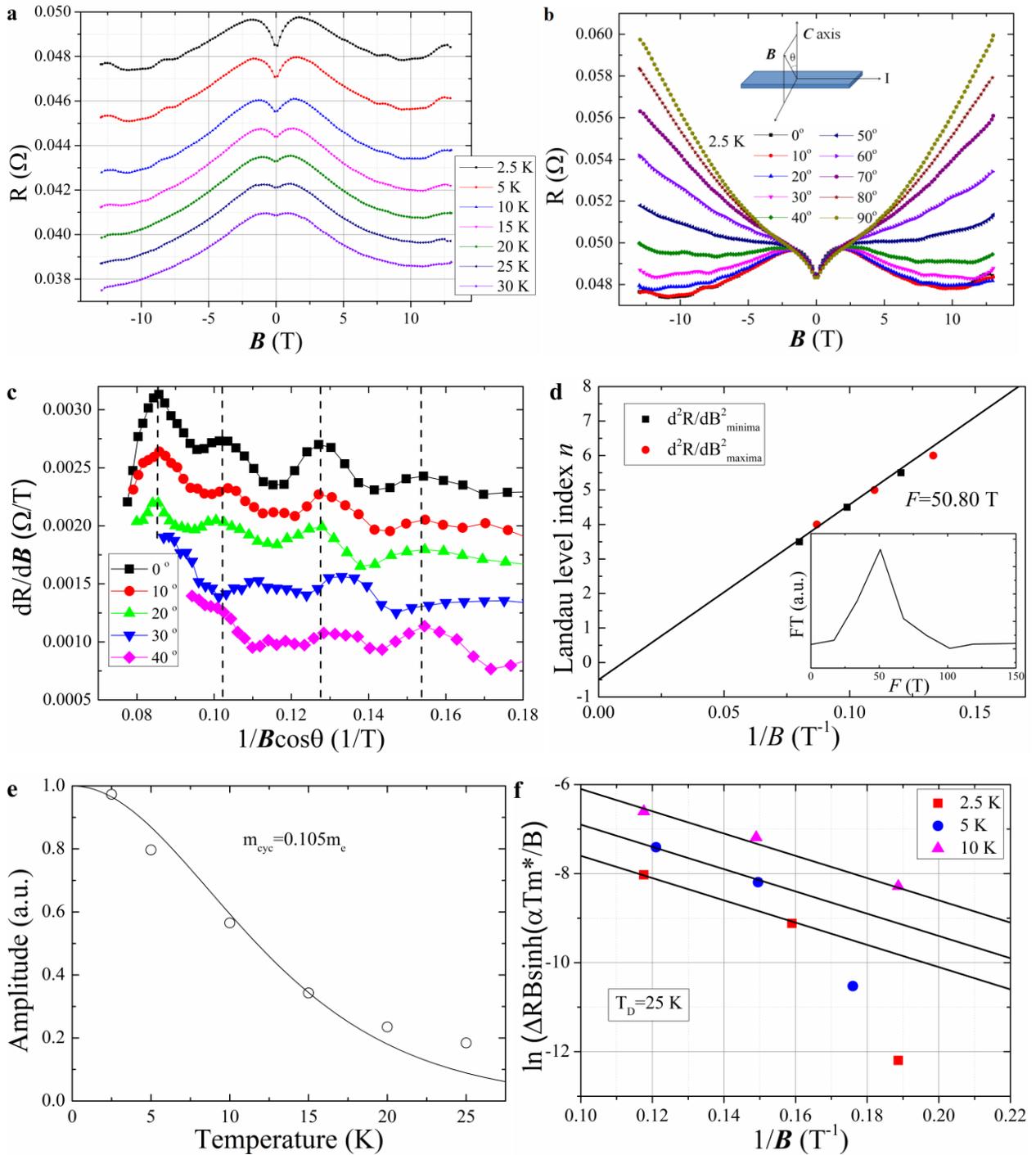

**Fig. 2** (a) Magnetoresistance up to 13 T was measured at various temperatures, and the weak anti-localization disappears with increasing temperature. (b) Angular dependence of magnetoresistance from 0º to 90º at 2.5 K. (c) d$R$/d$B$ is plotted as function of the inverse perpendicular component of the magnetic field, $\frac{1}{B\cos\theta}$, at various angles. (d) Landau level fan. Inset is the fast Fourier transform (FFT) of the oscillation, giving the oscillation frequency, $F$ = 50.80 T. (e) Temperature dependence of the oscillation amplitude, which yields the cyclotron effective mass, $m_{cyc}$ = 0.105 $m_e$. (f) Dingle plot of the oscillation $\Delta RB\sinh(\alpha Tm^*/B)$ versus $1/B$.

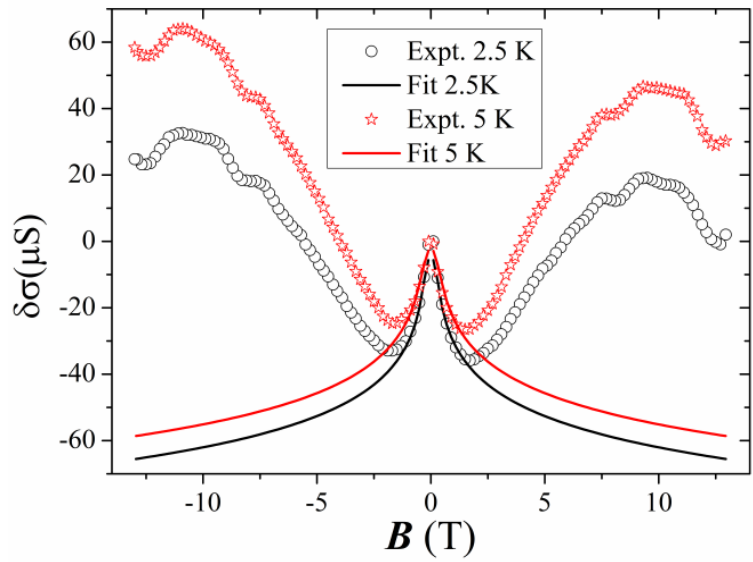

**Fig. 3** The magnetoconductance at 2.5 K and 5 K, which in low field, can be fitted by the HKN formula.

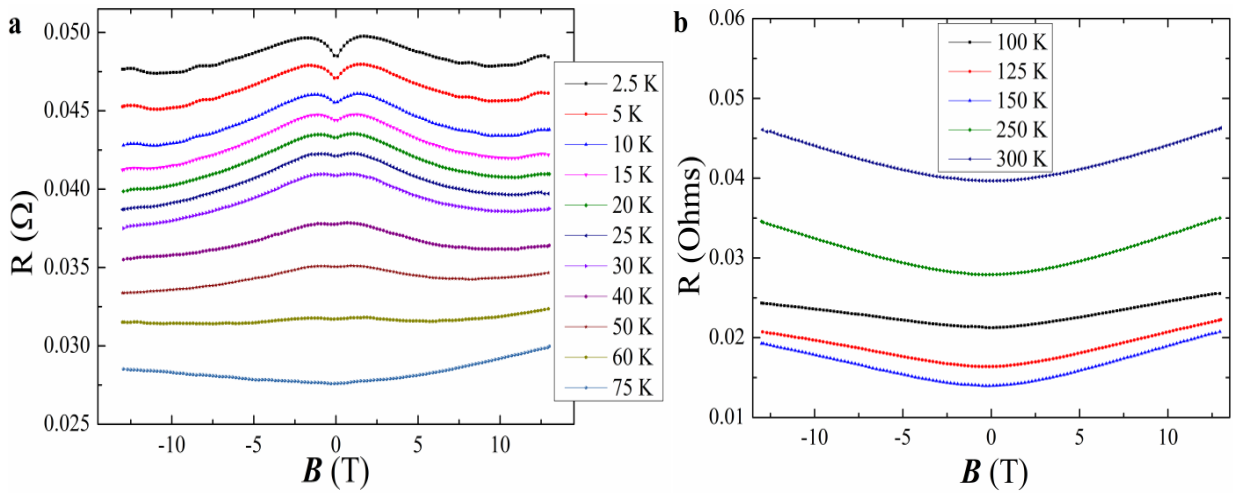

**Fig. 4** MR measured from 2.5 K to 75 K (a) and from 100 K to 300 K (b).